\documentclass[letter,12pt]{article}
\pdfoutput=1 

\usepackage{jheppub} 

\usepackage[utf8]{inputenc}

\usepackage[russian,english]{babel}
\usepackage[T1,T2A]{fontenc} 

\usepackage{amsmath}
\DeclareMathOperator{\Tr}{Tr}
\usepackage{ upgreek }

\def\bea{\begin{eqnarray}}
\def\eea{\end{eqnarray}}

\title{Exact generalized partition function of 2D CFTs at large central charge}

\author[a,b]{Anatoly Dymarsky} 
\author[b, c]{and Kirill Pavlenko}

\affiliation[a]{University of Kentucky,\\Lexington, KY, USA 40506\\}
\affiliation[b]{Skolkovo Institute of Science and Technology,\\Skolkovo Innovation Center, Moscow, Russia\\}
\affiliation[c]{Moscow Institute of Physics and Technology, Dolgoprudny 141700, Russia}
\emailAdd{a.dymarsky@uky.edu}
\emailAdd{kirill.pavlenko@skoltech.ru}

\abstract{We discuss generalized partition function of 2d CFTs decorated by higher qKdV charges on thermal cylinder.  We propose that in the large central charge limit qKdV charges factorize such that generalized  partition function can be rewritten in terms of auxiliary non-interacting bosons.
The explicit expression for the generalized free energy  is readily available in terms of the boson spectrum, which can be deduced from the conventional thermal expectation values of qKdV charges. In other words,  the picture of the auxiliary non-interacting bosons  allows extending thermal one-point functions to the full non-perturbative generalized partition function. We verify this conjecture for the first seven qKdV charges using recently obtained pertrubative results and find corresponding contributions to the auxiliary boson masses. We further extend these results by conjecturing the full spectrum of bosons and find an exact expression for the generalized partition function as a function of infinite tower of chemical potentials in the limit of  large central charge. } 

\notoc

\begin{document} 
\maketitle
\flushbottom

\section{Introduction}
\label{sec:intro}

Generalized partition function of 2d CFTs decorated by higher qKdV charges \cite{bazhanov1996integrable,bazhanov1997integrable,bazhanov1999integrable}, the so-called Generalized Gibbs Ensemble,
\bea
\label{Z}
Z=\Tr{\rm exp} \left\{-\sum\limits_{k=1}^\infty \mu_{2k-1}Q_{2k-1}\right\},\quad \mu_1\equiv \beta,\quad Q_1\equiv H,
\eea
has been in the focus of attention recently in the context of thermalization of large $c$ 2d conformal theories \cite{calabrese2007quantum,calabrese2011quantum,cardy2016quantum, de2016remarks,perez2016boundary,He2017vyf,basu2017thermality,He2017txy,lashkari2018universality,maloney2018generalized,maloney2018thermal,dymarsky2018generalized}. In this work we assume thermodynamic limit, when the size of the spatial circle goes to infinity $\ell\rightarrow \infty$ and \eqref{Z} describes theory on a thermal cylinder. 

In a recent work \cite{dymarsky2018generalized} we observed that in the large central charge limit first two non-trivial qKdV charges $Q_3,Q_5$ admit a simple structure. Schematically, 
\bea
\ell^{2k-1}Q_{2k-1}=L_0^k+\ell^{2k-1}\tilde{Q}_{2k-1}+O(c^{k-2}), \label{Q}
\eea
where we have neglected terms suppressed in the thermodynamic limit. 
Equation \eqref{Q} is an effective expansion in $1/c$,
first term 
$L_0^k$ contributes as $O(c^{k})$, while
$\tilde{Q}_{2k-1}$  contributes as $O(c^{k-1})$. 
Written  in the conventional basis of conformal theory (sets $\{m_i\}$, $m_1\geq m_2,\dots, \geq m_k$, are arranged in dominance order),
\bea
\label{basis}
|m_i,\Delta\rangle =L_{-m_1}\dots L_{-m_k}|\Delta\rangle, 
\eea
$L_0^k$ is diagonal and $\tilde{Q}_{2k-1}$ is lower-triangular. 
Here and below we assume that $\Delta$ scales  linearly with $c$.  
Remarkably, at first two leading orders in $1/c$ expansion the eigenvalues of $Q_{2k-1}$ (terms suppressed in the thermodynamic limit are neglected), 
\bea
\label{lambda}
\ell^{2k-1}Q_{2k-1}|\lambda \rangle= \lambda |\lambda \rangle, \quad \lambda=\Delta^k+\sum_{p=0}^{k-1} \sum_i m_i^{2p+1} c^{p}\Delta^{k-1-p}\, \xi_k^p+O(c^{k-2}),
\eea  
are linear in  the occupation numbers $n_r$,  provided the sets $\{m_i\}$ are rewritten in terms of the free boson representation,
\bea
\label{freeboson}
\sum_i m_i^p=\sum_r  r^p n_r.
\eea
The linearity of $\lambda$ in $n_r$ is crucial for what follows. Technically it is due the fact that \eqref{lambda} includes only  a single sum over $m_i$. If (\ref{lambda}) applies to all $Q_{2k-1}$, at first two orders in $1/c$ generalized partition function \eqref{Z} reduces to that one of non-interacting  auxiliary bosons with the spectrum given in terms of $\mu_{2k-1}$ and $\xi_k^p$. 

In principle the coefficients $\xi_k^p$ can be deduced directly from the explicit form of $Q_{2k-1}$ in terms of Virasoro generators $L_n$, as was done for $Q_3,Q_5$ in \cite{dymarsky2018generalized}. Extending this strategy to higher charges is difficult because their explicit form is not known and difficult to calculate. A much simpler way to obtain $\xi_k^p$ follows from the expression for thermal average of $Q_{2k-1}$ over a particular Verma module,
\bea
\langle Q_{2k-1} \rangle_{\beta,\Delta}=\Tr_\Delta (q^{L_0} Q_{2k-1}),\qquad q=e^{-\beta/\ell}, \label{tr}
\eea
where the sum in \eqref{tr} goes over all states of the form \eqref{basis} with a fixed $\Delta$. This one-point function was calculated recently  for the first seven  qKdV charges $Q_{2k-1}$, $k\leq 7$, in \cite{maloney2018thermal}. Using this result we confirm the proposed form of the eigenvalues \eqref{lambda} and obtain corresponding coefficients $\xi_k^p$. We notice these coefficients admit a simple form, which can be easily generalized to all $k$,
\bea
\xi_k^p= 24^{-p}{(2k-1)\Gamma(k+1)\Gamma(1/2)\over 2\,\Gamma(p+3/2)\Gamma(k-p)}.\label{xi}
\eea
Assuming that \eqref{lambda} and \eqref{xi} apply to all higher $Q_{2k-1}$,  generalized partition function at first two orders in $1/c$ expansion reduces to that one of non-interacting auxiliary bosons, yielding 
\bea
\label{Fotvet}
Z&=&e^{F},\ \ \, \quad\qquad \qquad  \qquad\qquad  \qquad F={\pi^2 \ell\over 6\, \beta^2}\left(c' f_0+f_1+ O(1/c')\right),\\
\nonumber f_0&=&\sum_{k=1}^\infty t_{2k-1}\sigma^k (2k-1),\qquad\qquad \, \sqrt{\sigma}=\sum^\infty_{k=1} t_{2k-1} \sigma^k k,\\ 
f_1&=&-{12\over \pi }\int_0^\infty d\kappa \log\left(1-e^{-2\pi \kappa \gamma}\right),\quad \gamma= \sum_{k=1}^\infty t_{2k-1} \sigma^{k-1} k(2k-1) {}_2F_1(1,1-k,3/2,-\kappa^2/\sigma), \nonumber \\
c'&=&c-1,\qquad t_{2k-1}=\left({\pi^2 c'\over 6\beta^2}\right)^{k-1}{\mu_{2k-1}\over \beta},\qquad t_1\equiv 1. \nonumber
\eea
An explicit expression for $\sigma(t)$ in terms of an infinite power series can be found in \eqref{sigmaseries}. The conjectural expression for $f_1$ is the main result of this paper. 

This paper is organized as follows: in the next section we discuss first seven qKdV charges $Q_{2k-1}$, $k\leq 7$, and verify they are consistent with \eqref{lambda}. We also calculate corresponding coefficients $\xi_k^p$ and conclude that  \eqref{xi} describes all of them. In section three we assume \eqref{lambda} and \eqref{xi} are valid beyond $k\leq 7$ for all $Q_{2k-1}$ and calculate generalized partition function \eqref{Fotvet}. The relation between $1/c$ and $1/c'$ expansion is discussed in the appendix.

\section{Thermal average of $Q_{2k-1}$}
In this section we discuss how the form of the eigenvalues \eqref{lambda} can be verified and the coefficients $\xi_k^p$ can be fixed from the explicit form of thermal one-point averages \eqref{tr} obtained in \cite{maloney2018thermal}.  Because of the lower-triangular form of $\tilde{Q}_{2k-1}$,  leading terms of $\lambda$   contribute to the thermal average \eqref{tr} as a linear combination of 
\bea
\label{sigma}
\sum_{\{m_i\}} \sum_i q^{\Delta +n} m_i^r=\sigma_r\chi,\qquad n\equiv \sum_i m_i, \qquad \chi={q^\Delta\over \prod_i (1-q^i)},
\eea
where $\sigma_k$ are related to Eisenstein series via
\bea
\sigma_p&=&\sum\limits_{k=0}^\infty {k^p q^k\over 1-q^k},\qquad  E_{2p}=1+{2\over \zeta(1-2p)}\sigma_{2p-1}.
\eea
In other words, to fix $\xi_k^p$ we need to find coefficients in front of $\sigma_{2p+1} c^p \Delta^{k-1-p}$.

\subsection{$Q_1$}
As a warm-up we start our analysis with 
\bea
\ell Q_1=L_0-{c\over 24}.
\eea
The constant term $-c/24$ does not contribute in the thermodynamic limit and therefore the structure \eqref{Q} is manifest with $\tilde{Q}_1=0$. The eigenvalues of $L_0=\Delta+n$, $n\equiv \sum_i m_i$, have the form \eqref{lambda} with $\xi_1^0=1$. Although this is straightforward we want to derive the same result in a slightly different way, 
\bea
\label{derivative}
\Tr_\Delta (q^{L_0} L_0)=\partial \chi=(\Delta +\sigma_1)\chi,\qquad \partial\equiv q\partial_q.
\eea 
Hence $\xi_1^0=1$  is simply the coefficient in front of $\sigma_1$.

\subsection{$Q_3$}
The explicit expression for $Q_3$ is bulky, 
\bea
\ell^3 Q_3=L_0^2 - \frac{c+2}{12} L_0  + \frac{c(5c+22)}{2880}+2\sum_{i=1}^\infty L_{-i} L_i,
\eea
but only first and last terms contribute in the thermodynamic limit yielding \eqref{Q} with $\ell^3 \tilde{Q}_3=2\sum_{i=1}^\infty L_{-i} L_i$. Thermal average \eqref{tr} can be calculated using trace cyclicity \cite{Apolo2015Q3}, yielding \cite{dymarsky2018generalized,maloney2018thermal}  
\bea
\ell^3\Tr_\Delta (q^{L_0} Q_3)=\left(D^2+{c\over 1440}E_4\right)\chi,
\eea
where here and below 
\bea
D^k=\left(\partial-{k-1\over 6} E_2\right)\left(\partial-{k-2\over 6} E_2\right)\dots \partial.
\eea
Leading term $\Delta^2$ follows from $\partial^2$. 
Using \eqref{derivative}, 
we calculate the  coefficients in front of $\Delta$ and $c$ 
\bea
\ell^3\Tr_\Delta (q^{L_0} Q_3)=\Delta^2+\Delta \left(6\sigma_1-{1\over 6}\right)+{c\over 6}\left(\sigma_3+{1\over 240}\right)+\partial \sigma_1. \label{q3}
\eea
To express $E_{2p}$ in terms of $\sigma_{2p-1}$ we need the numerical values of zeta-function, which we write down here for reader's convenience,
\bea
\nonumber
\zeta(-3)=-{1\over 12},\ \   \zeta(-3)={1\over 120},\ \  \zeta(-5)=-{1\over 252},\ \  \zeta(-7)={1\over 240},\\
 \zeta(-9)=-{1\over 132},\ \  \zeta(-11)=-{691\over 32760},\ \ \zeta(-13)=-{1\over 12}.\label{zeta}
\eea

We are only interested in the first two terms of $1/c$ expansion ($\Delta$ is assumed to scale linearly with $c$), hence the term $\partial \sigma_1$ from \eqref{q3} can be neglected. Next, we only consider the terms which contribute extensively in the thermodynamic limit $\ell\rightarrow \infty$. We assume that $\Delta$ scales as $\ell^2$ while the scaling of $\sigma_r\propto \ell^{r+1}$ follows from its explicit form. There is another more intuitive way to understand that directly from  \eqref{sigma}. Main contribution to the thermal average comes from the partitions $\{m_i\}$ which consist of approximately $n^{1/2}$ terms and each term $m_i \sim n^{1/2}$, while typical $n=\sum_i m_i$ scales as $\ell^2$. Keeping only the terms scaling as $\ell^4$ in \eqref{q3} we obtain
\bea
\ell^3\Tr_\Delta (q^{L_0} Q_3)=\Delta^2 +6\Delta\, \sigma_1+{c\over 6}\sigma_3+O(1/c), \label{otvet3}
\eea 
in full consistency with \eqref{lambda}. This result agrees with the calculation of \cite{dymarsky2018generalized}, which utilizes the explicit form of $Q_3$ in terms of Virasoro algebra generators. First term $L_0^2=(\Delta+n)^2$ yields $\Delta^2+2 \Delta n$, ($n^2$ can be neglected because it contributes as $c^0$), while the eigenvalue of $\ell^3\tilde{Q}_3={c\over 6}(\sum_i m_i^3-n)+4\Delta n$ completes it to (\ref{otvet3}), or (\ref{lambda}) with
$\xi_2^2=1/6$ and $\xi_2^1=4$.

\subsection{$Q_5$}
The calculation for $Q_3$ reveals the pattern how the terms of interest enter the full expression for the thermal average. The leading term $\Delta^k$  of the eigenvalue of $Q_{2k-1}$ follows from $D^k\chi$, as well as $\xi^0_{k-1}\Delta^{k-1}\sigma_1$. The term $\xi^1_{k-1}c\Delta^{k-2}\sigma_3$ follows from $cE_4 D^{k-2}\chi$, and so on. 
In case of $Q_5$ we have for the thermal average \cite{maloney2018thermal},
\bea
\ell^5\Tr_\Delta (q^{L_0} Q_5)=\left(D^3+{c+4\over 288}E_4 D-{c(c+14)\over 36288}E_6\right)\chi.
\eea
This yields in the  limit of interest
\bea
\ell^5\Tr_\Delta (q^{L_0} Q_5)=(\Delta^3+15\Delta^2 \sigma_1+{5\over 6}c \Delta \sigma_3+{1\over 72}c^2\sigma_5)\chi,
\eea
where the last term came from $c^2 E_6 D^{k-3}\chi$, $k=3$. This result is in full agreement with the explicit calculation of \cite{dymarsky2018generalized}.

\subsection{$Q_7$}
The original expression for $\Tr_\Delta (q^{L_0} Q_7)$ calculated in \cite{maloney2018thermal} is quadratic in $E_4$, but using the identify $E_4^2=E_8$ it can be written as follows 
\bea
\ell^7\Tr_\Delta (q^{L_0} Q_7)=\left(D^4+\frac{(7 c+64)}{720} E_4 D^2-\frac{c^2+24 c+74}{6480}E_6 D+\frac{c \left(c^2+\frac{103 c}{4}+175\right)}{518400}E_8\right)\chi. \nonumber
\eea
This immediately gives
\bea
\ell^7\Tr_\Delta (q^{L_0} Q_7)=\left(\Delta^4+28 \Delta^3\sigma_1+{7\over 3}c \Delta^2 \sigma_3+{7\over 90}c^2\Delta \sigma_5+{1\over 1080}c^3\sigma_7\right)\chi.
\eea
Corresponding values of $\xi_3^p$ are easy to obtain using numerical values  \eqref{zeta}.

\subsection{$Q_9$}
The expression for $Q_9$ is too bulky and here we only write relevant terms using $E_4^2=E_8$ and $E_4 E_6=E_{10}$,
\bea
\nonumber
\ell^9\Tr_\Delta (q^{L_0} Q_9)=\left(D^5+\left({7c\over 720}+O(c^0)\right)E_4 D^3
+\left(-{c^2\over 2016}+O(c^1)\right)E_2 D^2+ \right. \\ \left.\left(-{c^3\over 80640}+O(c^2)\right)E_8 D+\left(-{c^4\over 4790016}+O(c^3)\right)E_{10}\right)\chi.
\eea
Corresponding values of $\xi_4^p$ immediately follow from here. 

\subsection{$Q_{11}$, $Q_{13}$, and beyond}
Calculation of the  eigenvalues of $Q_{11}$ and $Q_{13}$ is completely analogous, but to rewrite the leading part of $\Tr_\Delta (q^{L_0} Q_{2k-1})$ as a linear combination of  $D^k$ and terms of the form  $c^{k-1-p}E_{2(k-p)}D^p$, $p=0,\dots,k-2$, we need to use the identities 
\bea
E_{12} = \frac{441}{691} E_4^3 + \frac{250}{691} E_6^2,\qquad E_{14}=E_4^2 E_6.
\eea
Resulting values of the coefficients $\xi_k^p$ for $k=1,\dots,7$, are summarized in the table below
\begin{eqnarray}
\xi_k^p=\left(
\begin{array}{ccccccc}
 1 &  &  &  &  &  &      \\
 6 & \frac{1}{6}  &  &  &  & &  \\
 15 & \frac{5}{6} & \frac{1}{72} &  &  & &  \\
 28 & \frac{7}{3} & \frac{7}{90} & \frac{1}{1080} &  &  & \\
  45 & 5  & \frac{1}{4} & \frac{1}{168}  & \frac{1}{18144}&   &  \\
66 & \frac{55}{6}  & \frac{11}{18} & \frac{11}{504}   & \frac{11}{27216} &  \frac{1}{326592}&   \\
 91  & \frac{91}{6} & \frac{91}{72}  & \frac{13}{216} &  \frac{13}{7776} &  \frac{13}{513216}  & \frac{1}{6158592} \\
\end{array}
\right),\quad  p=0,\dots,k-1.
\end{eqnarray}
These values can be concisely written as 
\bea
\xi_k^p= 24^{-p}{(2k-1)\Gamma(k+1)\Gamma(1/2)\over 2\,\Gamma(p+3/2)\Gamma(k-p)},\label{xi-new}
\eea
which extends this result for all $k$.

\section{Generalized partition function}
From now on we assume that \eqref{lambda} applies to all qKdV charges with the coefficients $\xi_k^p$ given by \eqref{xi-new}. Given that all $Q_{2k-1}$ mutually commute, the generalized partition function \eqref{Z} is given by the sum over primaries $\Delta$ and sets (Young tables) $\{m_i\}$, parameterizing descendants via \eqref{basis},
\bea
Z=\sum_\Delta \sum_{\{m_i\}} {\rm exp}\left(-\sum_{k=1}^\infty {\mu_{2k-1} \over \ell^{2k-1}}\left(\Delta^k+\sum_{p=0}^{k-1} \sum_i m_i^{2p+1} c^{p}\Delta^{k-1-p}\, \xi_k^p+O(c^{k-2})\right)\right).
\eea
At large central charge sum over $\Delta$ can be substituted by an integral 
\bea
\sum_\Delta \rightarrow \int d\Delta\, e^{\pi\sqrt{2c' \Delta/3}},\qquad c'\equiv c-1,
\eea
where the density of primaries follows from Cardy formula \cite{cardy1986operator,kraus2017cardy}. 
It is convenient to introduce $\sigma$ via
\bea
\Delta={c'\pi^2\ell^2\over 6\,\beta^2}\sigma.
\eea
So far we were discussing $1/c$ expansion, but the results look more elegant if we do an expansion in $1/c'$. Since at leading order $c=c'+O(1)$,  the structure of $\lambda$ remains the same: $\Delta^k$ contributes as $(c')^k$ while $c^p \Delta^{k-1-p}$ terms contribute as $(c')^{k-1}$. Going from the sets $\{m_i\}$ to free boson representation \eqref{freeboson}, the partition function reduces to that one of non-interacting auxiliary bosons
\bea
\label{gge}
&&Z(\beta,t)=\int d\sigma\, {\rm exp} \left\{{c'\pi^2 \ell\over 6\beta}\left(2\sqrt{\sigma}-\sum\limits_{k=1}^\infty t_{2k-1}\sigma^k\right)\right\}\sum_{n_1,n_2,\dots} e^{-\sum_{r=1}^\infty n_r M_r+O(1/c')},\quad \\
&&\log Z\equiv F={\pi^2 \ell\over 6\beta}\left(c' f_0(t)+f_1(t)+O(1/c')\right),\\
&&t_{2k-1}=\left({\pi^2 c'\over 6\beta^2}\right)^{k-1}{\mu_{2k-1}\over \beta},\qquad t_1\equiv 1,
\eea
where the spectrum of bosons is given by 
\bea
M_r&=&\sum_{k=1}^\infty t_{2k-1} \sigma^{k-1}\sum_{p=0}^{k-1} \xi_k^p \left({6\over \pi^2 \sigma}\right)^p \left({\beta r\over \ell}\right)^{2p+1}=\\ 
&&{\beta r\over \ell} \sum_{k=1}^\infty t_{2k-1} \sigma^{k-1} k(2k-1)\, {}_2F_1\left(1,1-k,3/2,-{1\over \sigma}\left({\beta r\over 2\pi \ell}\right)^2\right).
\eea
In \eqref{gge} we write the partition function as a function of $\beta,t_{2k-1}$. For the given fixed $\beta,t_{2k-1}$  the terms contributing as $(c')^{k-2}$ to eigenvalues of $Q_{2k-1}$ contribute to free energy as $1/c'$.  Our scope is to calculate free energy up to the first two orders in $1/c'$ expansion, i.e.~only keep the terms which survive in the $c'\rightarrow \infty$ limit. Hence $O(1/c')$ terms can be neglected. 

Up to $1/c'$ corrections the value of $\sigma$ is determined via saddle point approximation of 
\bea
Z_0(\beta,t)={\rm exp} \left\{{c'\pi^2 \ell\over 6\beta}f_0\right\}=\int d\sigma\, {\rm exp} \left\{{c'\pi^2 \ell\over 6\beta}\left(2\sqrt{\sigma}-\sum\limits_{k=1}^\infty t_{2k-1}\sigma^k\right)\right\},
\eea
while the remaining sum over the boson occupation numbers $n_r$ in \eqref{gge}  ``takes'' saddle point value of $\sigma$ as an input. The saddle point equation 
\bea
\sqrt{\sigma}&=&\sum_{k=1}^\infty t_{2k-1} \sigma^k k,
\eea
can be solved in terms of an infinite series 
\bea
\sigma&=&1+\sum_{n=1}^\infty \sum_{k_1,\dots,k_n=2}^\infty  2{(-1)^n\over n!} {(2K-n+1)!\over (2K-2n+2)!} \prod_{i=1}^n k_i\, t_{2k_i-1},\quad
K\equiv\sum_{i}k_i,
\label{sigmaseries}
\eea
yielding (expansion \eqref{expansion} was found in \cite{maloney2018generalized}),
\bea
f_0&=&\sum_{k=1}^\infty t_{2k-1} \sigma^k (2k-1), \label{f0}\\
f_0&=&1+\sum_{n=1}^\infty \sum_{k_1,\dots,k_n=2}^\infty  2{(-1)^n\over n!} {(2K-n)!\over (2K-2n+2)!} \prod_{i=1}^n k_i\, t_{2k_i-1},\quad
K\equiv\sum_{i}k_i. \label{expansion}
\eea
With $\sigma$ being fixed, the remaining part of the partition function describes some auxiliary non-interacting bosons
\bea
{\pi^2 \ell\over 6\beta}f_1=\log \sum_{n_1,n_2,\dots} e^{-\sum_{r=1}^\infty n_r M_r}=-\sum_{r=1}^\infty \log\left(1-e^{-M_r}\right).
\eea 
In the thermodynamic limit $\ell\rightarrow \infty$ summation over $r$ can be substituted by integration (Thomas–Fermi approximation), yielding \eqref{Fotvet}.

\section{Discussion}
In this paper we have conjectured leading form of the spectrum of qKdV charges in $1/c$ expansion and verified it using recently obtained thermal averages for the first seven qKdV charges \cite{maloney2018thermal}. Using the conjectural form of the eigenvalues we have rewritten generalized partition function of 2d CFTs at large central charge in terms of non-interacting auxiliary bosons. The result of our calculation is the explicit form of the extensive part of free energy, exact up to $1/c$ corrections \eqref{Fotvet}.
We postpone discussing physical implications of our fundings until a future work. 

\acknowledgments
AD is supported by the National Science Foundation under Grant No.~PHY-1720374. AD is grateful to KITP for hospitality, where this work was initiated. The research at KITP was supported in part by the National Science Foundation under Grant No.~NSF PHY-1748958.


\appendix
\section{$1/c$ versus $1/c'$ expansion}
In a recent work \cite{dymarsky2018generalized} we were discussing free energy in $1/c$ expansion 
\bea
F={\pi^2\ell\over 6\beta}\left(c\tilde{f}_0(\tilde{t})+\tilde{f}_1(\tilde{t})+O(1/c)\right),
\eea
using variables 
\bea
\tilde{t}_{2k-1}=\left({\pi^2 c\over 6\beta^2}\right)^{k-1}{\mu_{2k-1}\over \beta}.
\eea
In this paper we used on $1/c'$ expansion 
\bea
F={\pi^2\ell\over 6\beta}\left(c'{f}_0({t})+{f}_1({t})+O(1/c')\right),
\eea
and the variables 
\bea
{t}_{2k-1}=\left({\pi^2 c'\over 6\beta^2}\right)^{k-1}{\mu_{2k-1}\over \beta}.
\eea
Here we outline the relation between these two expansion schemes. Using 
\bea
t_{2k-1}=\tilde{t}_{2k-1}\left(1-{1\over c}\right)^{k-1}
\eea
we readily find
\bea
\tilde{f}_0(t)=f_0(t),
\eea
and 
\bea
\tilde{f}_1(t)=-f_0(t)-\sum_{k=1}^\infty(k-1)t_{2k-1}{\partial f_0(t)\over \partial t_{2k-1}}+f_1(t).
\eea
Using the explicit form of $f_0$, \eqref{f0}, this can be simplified as
\bea
\tilde{f}_1(t)=-\sqrt{\sigma(t)}+f_1(t).
\eea
A comparison of $f_1$ from \eqref{Fotvet} with the equations (2.43), (2.52) of \cite{dymarsky2018generalized} confirms this result.

\bibliographystyle{JHEP}
\bibliography{GGE2}
\end{document}